\documentclass[final]{aipproc}
\pdfoutput=1
\layoutstyle{6x9}

\usepackage{journal_shortcuts}
\begin{document}

\newcommand{\degree}{^o}
\newcommand{\K}{\,\textrm{K}}
\newcommand{\Kpc}{\,\textrm{kpc}}
\newcommand{\Mpc}{\,\textrm{Mpc}}
\newcommand{\PC}{\,\textrm{pc}}
\newcommand{\Yr}{\,\textrm{yr}}
\newcommand{\CM}{\,\textrm{cm}}
\newcommand{\Myr}{\,\textrm{Myr}}
\newcommand{\Gyr}{\,\textrm{Gyr}}
\newcommand{\Kms}{\,\textrm{km}\,\textrm{s}^{-1}}
\newcommand{\Cm}{\,\textrm{cm}}
\newcommand{\Erg}{\,\textrm{erg}}
\newcommand{\ccm}{\,\textrm{cm}^{-3}}
\newcommand{\gccm}{\,\textrm{g}\,\textrm{cm}^{-3}}
\newcommand{\Presunit}{\,\textrm{erg}\,\textrm{cm}^{-3}}
\newcommand{\MicroG}{\,\mu\textrm{G}}

\newcommand{\Sun}{_{\sun}}
\newcommand{\Rot}{_{\mathrm{rot}}}
\newcommand{\Max}{_{\mathrm{max}}}
\newcommand{\Min}{_{\mathrm{min}}}
\newcommand{\Gal}{_{\mathrm{gal}}}
\newcommand{\ICM}{_{\mathrm{ICM}}}
\newcommand{\DM}{_{\mathrm{DM}}}
\newcommand{\Gas}{_{\mathrm{gas}}}
\newcommand{\Stars}{_*}
\newcommand{\Bulge}{_{\mathrm{bulge}}}
\newcommand{\Ram}{_{\mathrm{ram}}}
\newcommand{\KH}{_{\mathrm{KH}}}
\newcommand{\Grav}{_{\mathrm{grav}}}

\title{The dynamical intracluster medium:
a combined approach of observations and simulations}


\author{Elke Roediger}{
  address={Jacobs University Bremen, PO Box 750\,561, 28725 Bremen, Germany}
}

\author{Marcus Br\"uggen}{
  address={Jacobs University Bremen, PO Box 750\,561, 28725 Bremen, Germany}
}

\author{Aurora Simionescu}{
  address={Max-Planck-Institute for Extraterrestrial Physics, Giessenbachstr, 85748, Garching, Germany} 
}

\author{Hans B\"ohringer}{
  address={Max-Planck-Institute for Extraterrestrial Physics, Giessenbachstr, 85748, Garching, Germany}
  }

\author{Sebastian Heinz}{
  address={Department of Astronomy, University of Wisconsin, 475 N Charter Street Madison, WI 53706, USA}
}

\begin{abstract}
Current high resolution observations of galaxy clusters reveal a dynamical intracluster medium (ICM).  The wealth of
structures includes signatures of interactions between active galactic nuclei (AGN) and the ICM, such as cavities and
shocks, as well as signatures of bulk motions, e.g.~cold fronts. 
Aiming at understanding the physics of the ICM, we study individual clusters by both, deep high resolution observations and numerical simulations which include processes suspected to be at work, and aim at reproducing the observed properties. By comparing observations and simulations in detail, we gain deeper insights into cluster properties and processes. 
Here
we present two examples of our approach: 
the large-scale shock in the Hydra A cluster, and sloshing cold fronts. 
\end{abstract}

\maketitle


\section{The large-scale shock in Hydra A}

\begin{figure}
  \includegraphics[width=.42\textwidth]{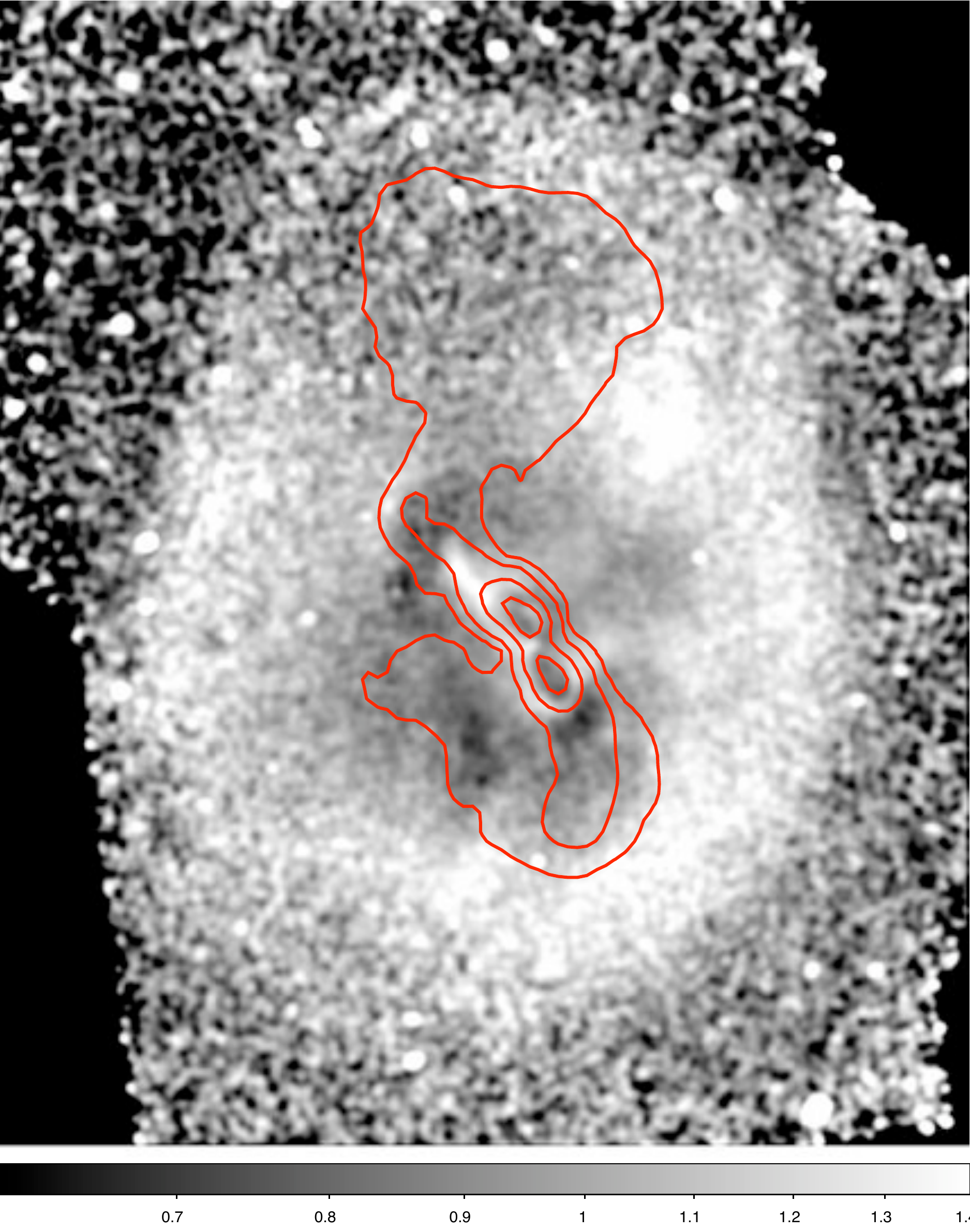}
  \includegraphics[height=.57\textwidth]{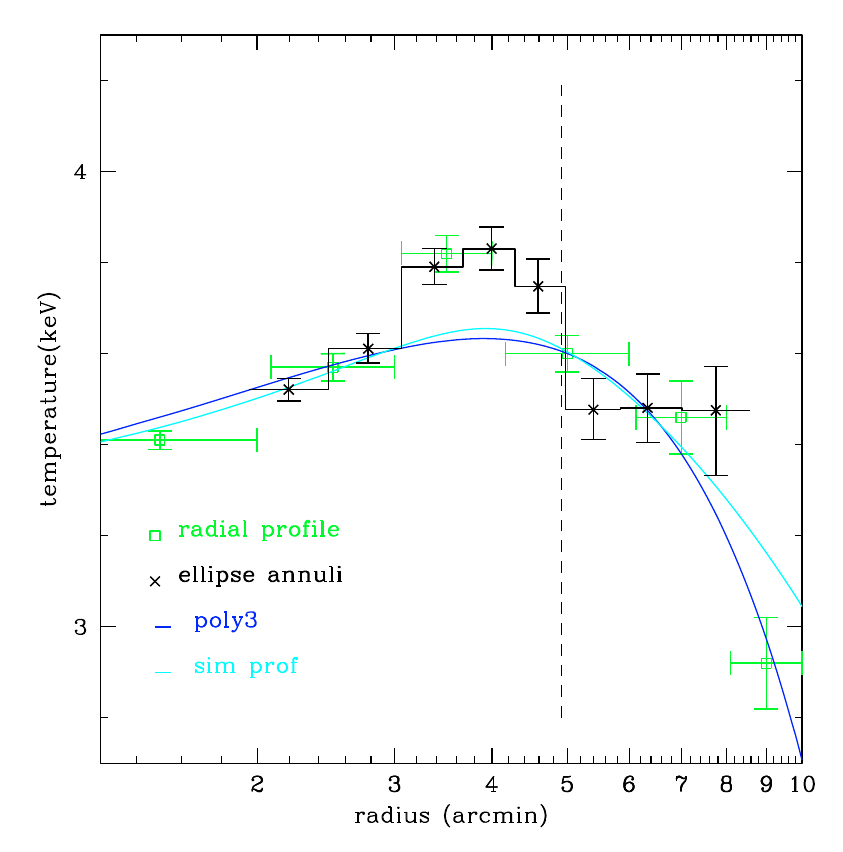}
    \caption{Hydra A observed. Left: residual X-ray image (background, A.~Simionescu) and radio contours (\citealt{Lane04}). 
      Right: Temperature profile in spherical or elliptical annuli, from \citealt{Simionescu09}. The vertical dashed line marks the shock. 
  }
  \label{fig:HydraObs}
\end{figure}

The galaxy cluster Hydra A is well-known for its AGN activity, evident not only in a pair, but a whole set of X-ray cavities (\citealt{McNamara00,Nulsen02,Nulsen05,Wise07}). Moreover, \citet{Nulsen05} and \citet{Wise07} detected a large-scale ($\sim 400 \Kpc$) surface brightness discontiunuity encircling the largest pair of cavities (Fig.~\ref{fig:HydraObs}, left panel). 
These authors interpreted the surface discontinuity as a shock caused by the same AGN outbreak that created the largest pair of  X-ray cavities. However, the temperature jump that should be associated with such a shock was not yet detected. 

\subsection{Observations}

We analyzed a new deep XMM-Newton observation of Hydra A, focusing on the large-scale shock 
described above (\citealt{Simionescu09}). 
The shock front can be seen, both, in the pressure map and in temperature 
profiles (Fig.~\ref{fig:HydraObs}, right panel). Thus, we can confirm the shock nature of the surface brightness discontinuity. The Mach number of the shock inferred in several sectors is $\sim 1.3$. The shape of the shock can be approximated with an ellipse centered $\sim 70 \Kpc$ towards the NE from the cluster center. In addition to this offset, the northern radio lobes appear larger than the southern ones.

\subsection{Simulations}
If the AGN interacts with a dynamical ICM, e.g. a large-scale bulk flow, the resulting structure - the combination of  radio lobes, cavities, and shock - could display the observed asymmetry and offset. We explore this scenario by means of 3D hydrodynamical simulations  (\citealt{Simionescu09}). The shock is produced by a symmetrical pair of AGN jets launched in a spherical galaxy cluster. The simulation successfully reproduces the size, ellipticity, and average Mach number of the observed shock front. The predicted age of the shock is 160 Myr and the total input energy $3 \times 10^{61} \Erg$. To match the observed 70 kpc offset of the shock ellipse  by large-scale coherent motions, these would need to have a high velocity of $670 \Kms$.
%
\begin{figure}
  \includegraphics[height=.3\textheight]{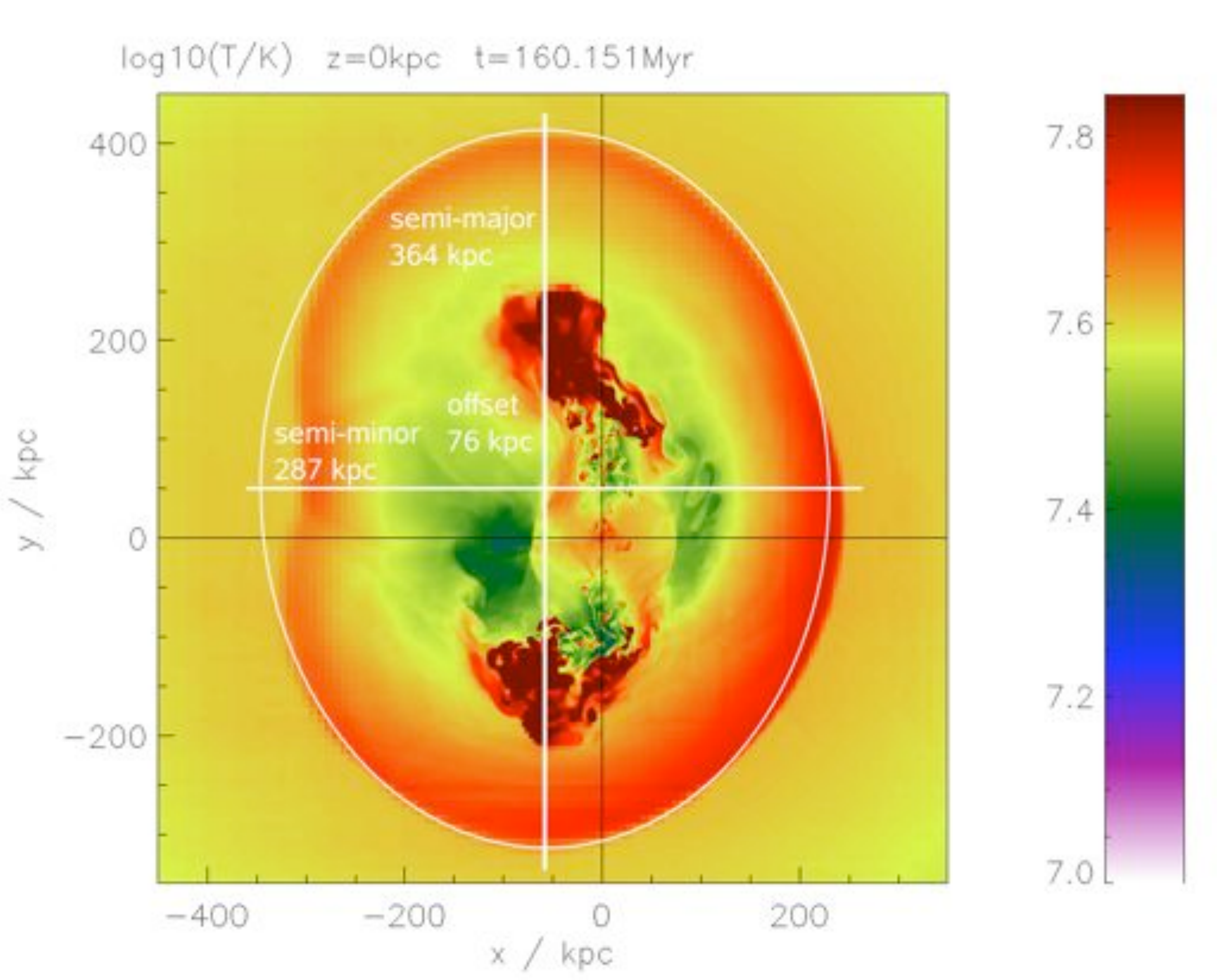}
  \caption{Hydra A simulated: Slice through the computational grid, color-coding shows temperature (from \citealt{Simionescu09}).} \label{fig:HySim}
\end{figure}
%
Although this scenario can explain the overall structure (Fig.~\ref{fig:HySim}), the required velocity is rather high. Alternative scenarios include a motion of the AGN w.r.t. the ICM, a non-spherical cluster structure, or a combination of different effects.

\section{Cold fronts}

In recent years, cold fronts have been detected in many clusters (see review by \citealt{Markevitch07} and references therein). They become evident as surface brightness discontinuities, where, in contrast to shocks, the brighter side is the cooler one. The discontinuities in surface brightness, temperature and density tend to very sharp. The pressure is approximately continuous over the front. Besides cold fronts associated with obviously merging clusters, another, somewhat weaker, variety has been detected to be "wrapped" around the cores of many otherwise relaxed and cool core clusters (\citealt{Ghizzardi06,Ghizzardi07}). The best-fit scenario available so far to explain these cold fronts is the idea of ICM sloshing, where a subcluster or, in best case, a dark matter only substructure passes near the cluster core, offsets the ICM, which then sloshes inside the cluster potential (\citealt{Markevitch01,Ascasibar06}). Recently, also discontinuities in metalicity associated with such cold fronts have been observed (e.g.~\citealt{Fabian05centaurus,Dupke07}), providing additional constraints on models. 

\begin{figure}
  \includegraphics[width=.49\textwidth]{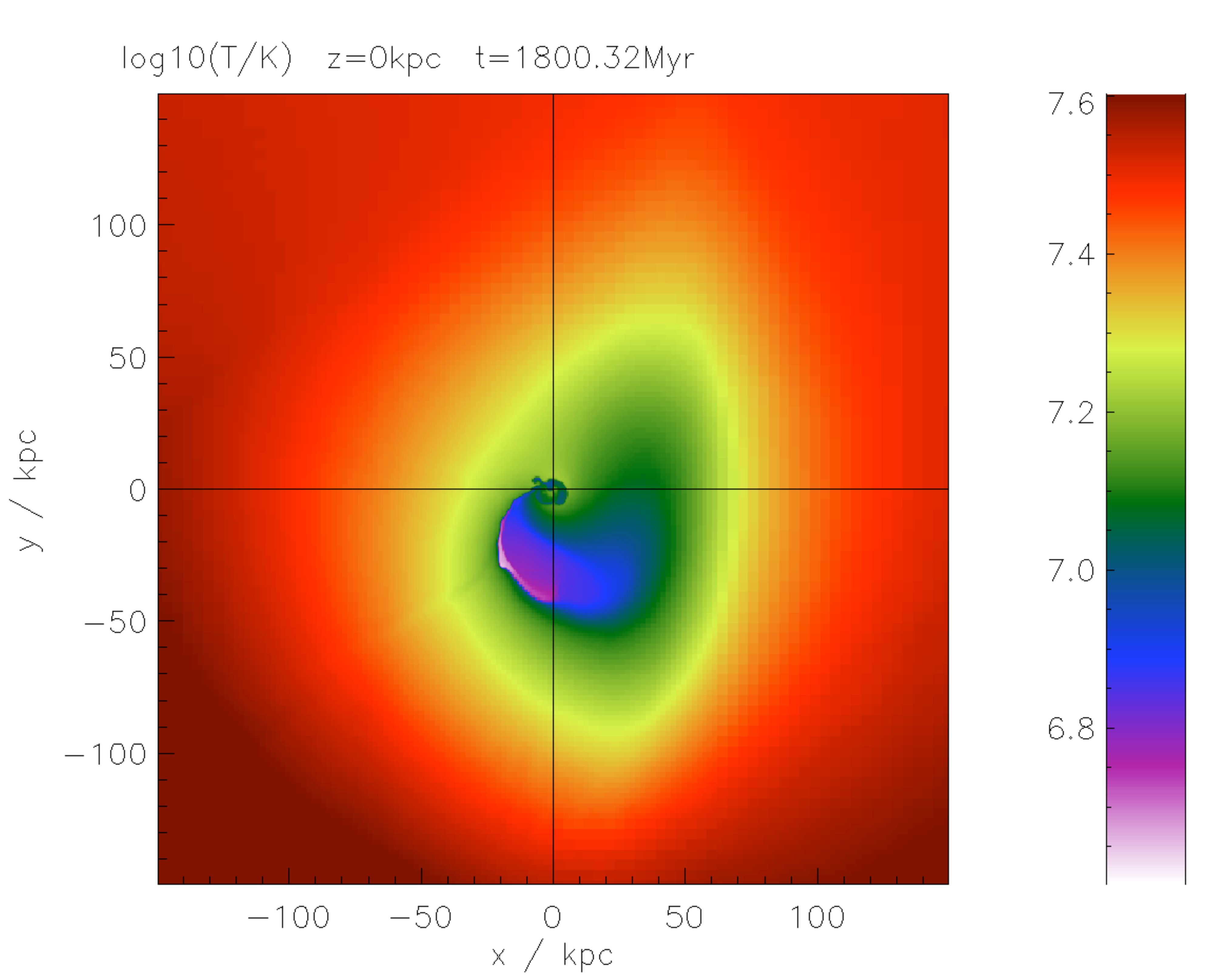}
  \includegraphics[width=.49\textwidth]{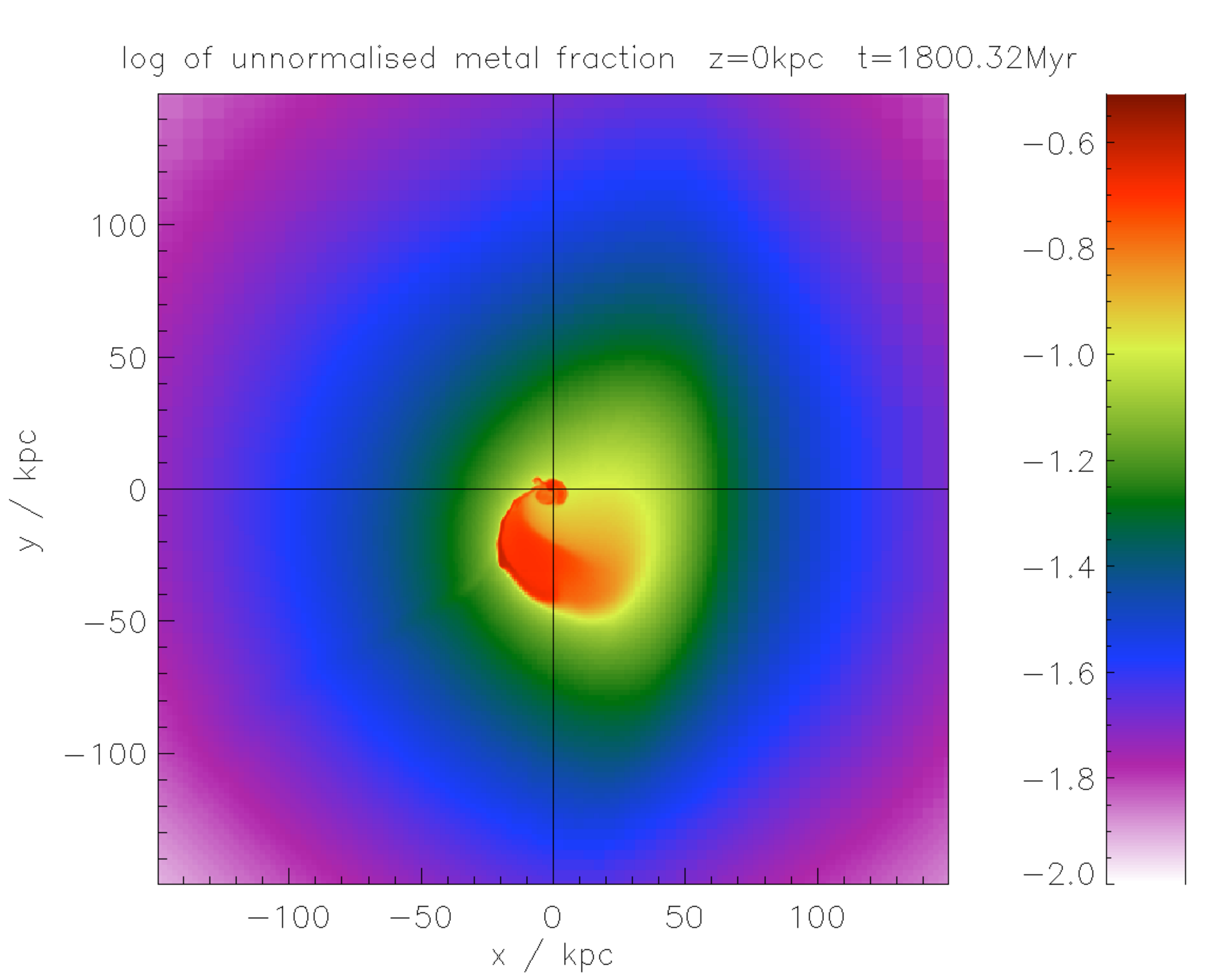}
  \caption{Cold front in a Virgo-like cluster: slice through the computational grid, color-coding shows temperature (left) and unnormalised metal fraction (right).} \label{fig:cf}
\end{figure}

We have run high resolution 3D hydrodynamical simulations of the sloshing scenario (Roediger et al. in prep.) which also trace the metal distribution in the ICM. These simulations can produce both, temperature and metalicity discontinuities (Fig.~\ref{fig:cf}). 
%
%
We apply this model to observed clusters to verify or disprove the sloshing scenario for them. 
For this purpose, we model ICM sloshing in both, spherical and elliptical clusters, and study the influence of cluster shape on the structure of the cold fronts. 
%
\begin{figure}
  \includegraphics[width=.47\textwidth]{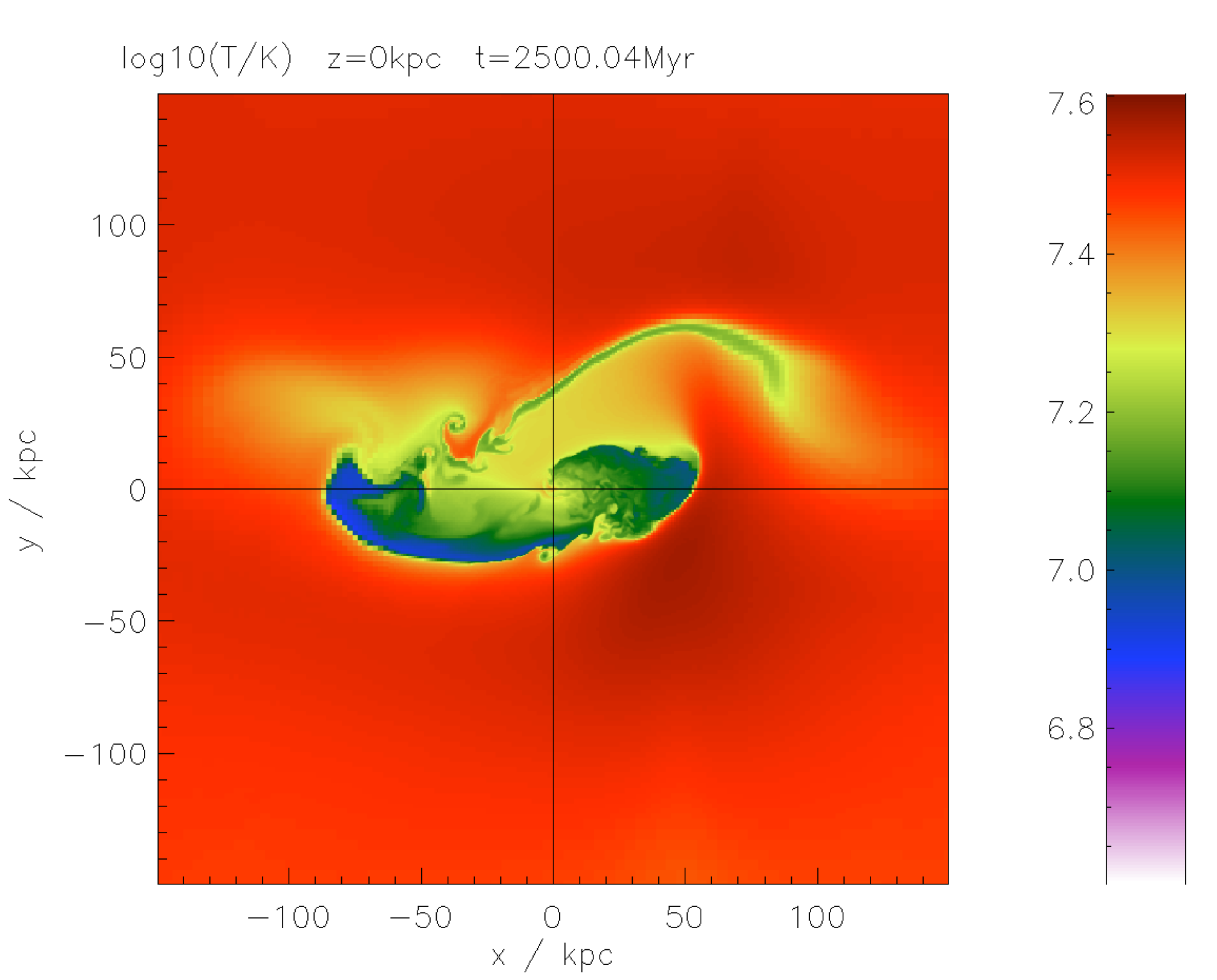}
  \caption{Cold front in an elliptical Virgo-like cluster: slice through the computational grid, color-coding shows temperature. Note the elongated shape of the cold front.}
\end{figure}


\begin{theacknowledgments}
We acknowledge the support of the Priority
Programme ``Witnesses of Cosmic History'' of the DFG (German Research Foundation) and the supercomputing  
grant NIC
2195 at the John-Neumann Institut at the Forschungszentrum J\"ulich.
The results presented were produced using the FLASH code, a product  
of the DOE
ASC/Alliances-funded Center for Astrophysical Thermonuclear Flashes  
at the
University of Chicago.
\end{theacknowledgments}

\bibliographystyle{aipproc}   

\bibliography{newbib_clean}

\end{document}